\documentclass[pra,aps,groupaddress,showpacs,prd,twocolumn]{revtex4}
\usepackage{epsfig,amsmath,graphicx,amssymb}
\usepackage[usenames]{color}
\def\be{\begin{equation}}
\def\ee{\end{equation}}
\def\bea{\begin{eqnarray}}
\def\eea{\end{eqnarray}}
\def\bes{\begin{subequations}}
\def\ees{\end{subequations}}
\newcommand{\hpsi}{ \hat{\psi}}

\begin{document}
%%%%%%%%%%%%%%%%%%%%%%%%%%%%%%%%%%%%%%%%%%%%%%%%%%%%%%%%%%%%%%%%%%%%%%%%%%%%%%%%%
\title{All-optical steering of light via spatial Bloch oscillations in a
gas of three-level atoms}

\author{Chao Hang$^{1,2}$ and V. V. Konotop$^{1,3}$}
\affiliation{ $^1$Centro de F\'isica Te\'orica e Computacional,
Universidade de Lisboa, Complexo Interdisciplinar, Avenida Professor
Gama Pinto 2, Lisboa 1649-003, Portugal
\\
$^2$Department of Physics, East China Normal University, Shanghai
200062, China
\\
$^3$Departamento de F\'isica, Faculdade de Ci\^encias, Universidade
de Lisboa, Campo Grande, Edif\'icio C8, Piso 6, Lisboa 1749-016, Portugal
}

%\date{\today}

%%%%%%%%%%%%%%%%%%%%%%%%%%%%%%%%%%%

\begin{abstract}

A standing-wave control field applied to a three-level atomic medium
in a planar hollow-core photonic crystal waveguide creates periodic
variations of linear and nonlinear refractive indexes of the medium.
This property can be used for efficient steering of light. In this
work we study, both analytically and numerically, the dynamics of
probe optical beams in such structures.  By properly designing the
spatial dependence of the nonlinearity it is possible to induce
long-living Bloch oscillations of spatial gap solitons, thus
providing desirable change in direction of the beam propagation
without inducing appreciable diffraction. Due to the significant
enhancement of the nonlinearity, such self-focusing of
the probe beam can be reached at extremely weak light intensities.
\end{abstract}

\maketitle

%%%%%%%%%%%%%%%%%%%%%%%%%%%%%%%%%%%%%%%%%%%%%%%%%%%%%%%%%%%%%%%%%
\section{Introduction}

Beam steering is one of the most important technologies in
the modern optics due to its numerous applications in  such fields
as optical imaging, laser machining, and free space communication.
Various physical mechanisms have been explored for
deflection of beams by inducing refractive index gradient \cite{GIL}. They
include mechanical motion \cite{CMFT}, thermal gradient
\cite{JABF}, the acousto-optical interaction \cite{DP}, and the
electro-optic effect \cite{SLB}. Fast beam steering in photonic
crystals \cite{KKTNTSK} and phased arrays \cite{VRHS} were also
proposed.

The direction of the light propagation can also be changed by
another beam of light through interaction with matter. In particular, a medium
exhibiting electromagnetically induced transparency (EIT) \cite{FIM}
can provide large probe-beam deflection since the refractive index
changes significantly near the transparency center \cite{HFI}. Among
the related studies we mention the refractive index measurements by
probe-beam \cite{PJAH}, investigation of slow light
deflection by magneto-optically controlled atomic
media~\cite{ZZWSS}, and exploring a scheme of all-optical beam
steering~\cite{SRZ}. Yet another example is the electromagnetically
induced waveguiding which uses the control field as a waveguide to
confine the probe field~\cite{MSFSD}.

However, the schemes proposed in the most of the previous studies are
restricted to the linear regime. They usually result in a spread of
the probe pulse because the refractive index gradient depends on
both the probe frequency and the spatial coordinates. Fortunately,
such spread can be suppressed by the enhancement of the wave
localization through the formation of solitons. Large
intrinsic nonlinearity in an EIT-based media  allows for
existence of probe beam solitons with extremely weak light
intensities~\cite{Schmidt}. More specifically, ultraslow solitons \cite{USOS},
spatial solitons \cite{SS}, and gap solitons~\cite{HKH} at low light intensity can exist in such kind of media.

In this work, we propose a scheme to achieve efficient all-optical
steering of light in a resonant three-level atomic system under EIT
regime. The scheme is based on the phenomenon of nonlinear
long-living Bloch oscillations (BOs) of gap solitons, recently
reported in Ref.~\cite{SKB}. The system at hand is a gas of
$\Lambda$-atoms loaded in a planar waveguide created by two photonic
crystals. The control field used in the system is a standing-wave
laser beam that originates a linear force, as well as
\textit{linear} and \textit{nonlinear} optical lattices (OLs)
affecting a weak probe beam. OLs in such a system are flexible: their
parameters can be adjusted either by changing the geometry
and/or intensity of the control field, or by varying one-- and/or
two--photon detunings.

Briefly, the mechanism of the steering consists of two ingredients: the
spatially periodic linear force induces dynamics of the probe
beam in the transverse direction, while the combined effect of the
lattices controls the direction of the beam propagation introducing
desirable and controllable deviations. More specifically, the linear
OL provides the band structure necessary for existence of the Bloch
states, while the nonlinear lattice is used to control the stability
properties of the Bloch states necessary for existence of gap
solitons at both edges of each band~\cite{SKB}. Due to the
enhancement of the Kerr nonlinearity induced by the EIT mechanism,
stable spatial gap solitons, representing the beam channels, can be
formed even subject to extremely weak probe light intensity and
subsequently strongly deflected without undergoing appreciable
deformations and attenuations.

The paper is organized as follows. In the next section, the model is
introduced.  In Sec. III, the nonlinear equation governing
the evolution of the probe field amplitude is derived. We show that stable
spatial gap solitons can exist under properly chosen parameters of
the control field. We also show how one can implement an efficient
all-optical steering of the gap solitons. All results in this work
are obtained under a set of experimentally feasible parameters. The
outcomes  are summarized in the Conclusion.

%%%%%%%%%%%%%%%%%%%%%%%%%%%%%%%%%%%%%%%%%%%%%%%%%%%%%%%%%%%%%%%%%
\section{The model}

\subsection{Preliminary arguments}

\begin{figure}
\centering
\includegraphics[scale=0.6]{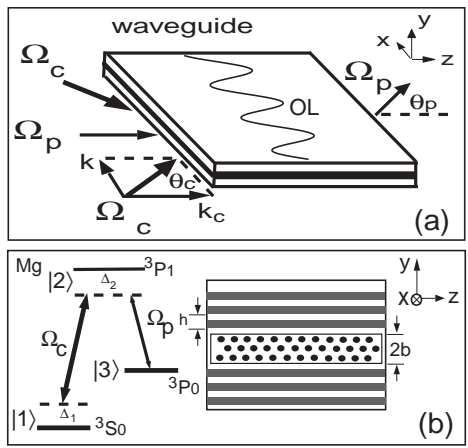}
\caption{(a) The  geometry considered in the present work:
 the  direction of propagation of the probe
beam ($z$-direction) is orthogonal to the OL axis ($x$-direction).
$\theta_c=\arctan(k/k_c)$ is the half angle between the two input
control beams and $\theta_p$ is the output  angle of the probe beam.
(b) Left panel: $\Lambda$-type atomic system interacting with two
laser beams. $\Omega_p$ and $\Omega_c$ are respectively the Rabi
frequencies of the probe and control fields. $\Delta_1$ and
$\Delta_2$ are respectively the two-photon and the one-photon
detunings. Right panel: The waveguide cross section. The dotted
region denotes the core filled in with the atomic gas and limited by
the Bragg mirrors. $2b$, $h_a$, and $h_b$ are the thicknesses of the
core and different materials of the Bragg mirrors, respectively. }
\label{fig1}
\end{figure}

We consider a cold gas of lifetime-broadened
$\Lambda$-type atoms, loaded into an anti-relaxation-coated planar
hollow-core photonic crystal waveguide, schematically shown in
Fig.~\ref{fig1}. (It is relevant to note that both room-temperature
and ultracold atoms have been successfully loaded into hollow-core
photonic crystal fibers~\cite{Ghosh}, and have further been used to
study EIT~\cite{GSOG} and all-optical switching~\cite{Bajcsy}). As
in any $\Lambda$-system, the atoms can populate three states: the
ground state $|1\rangle$, the excited state $|2\rangle$, and the
low-energy state $|3\rangle$ [see Fig.~\ref{fig1} (b)]. The transitions
between the states $|1\rangle$ and $|3\rangle$ are forbidden.
%===========================fig1===============================%
%===========================fig1===============================%
%

For the particular choice of the atomic gas and geometry of the
system, we take into account the two facts. First, in order to
contribute to the nonlinear polarization  the matrix elements of the
transition between the lower states and the excited state must be
nonzero, and hence the population of the excited state must also be
nonzero (see e.g.~\cite{HKH}).  Thus, one has to explore a gas with
weak atomic losses due to spontaneous emission from the excited
state.  A particular system having such properties
is the laser-cooled alkaline earth metal atoms, such as strontium
atoms~\cite{Traverso} where the atomic states $|1\rangle
$, $|2\rangle $, and $|3\rangle $ can be chosen as $^{1}\!S_0$, and
$^{3}\!P_1$, and $ ^{3}\!P_0$, respectively. One of the advantages
of such a system is that the atoms possess a long life time even in their
excited states: the life time of the state $|2\rangle$ is about
$21.3$ $\mu$s corresponding to the decay rate $47$ kHz. Therefore,
in what follows we  use this system for the sake of  the numerical studies
(see Figs.~\ref{fig2},~\ref{fig3}, and~\ref{fig4}).

The second fact to be taken into account is the smallest possible
size of the system. Our aim is to execute effective steering of
light using OLs. In order to make the periodicity appreciable in a
finite structure, one has to require the system to have the width of
at least ten lattice periods or more. Hence, the desirable OL should
be created by the beam with the highest possible frequency. In our
case, there are only two electromagnetic waves in the system, which
provide the transitions $1\leftrightarrow2$ and $3\leftrightarrow2$,
respectively. Thus we are interested in the geometry where the
strong control field connects the ground $|1\rangle$ and the excited
$|2\rangle$ states, while the probe weak wave links the states
$|1\rangle$ and $|3\rangle$.  In the case of the
strontium atoms~\cite{Traverso} the wavelength of the probe laser
beam is $\lambda_{p}=52$ $\mu$m while the length of the coupling
field is $\lambda_{c}=689$ nm. Requiring the modulation of the
nonlinear polarization to have at least 20 lattice periods,
we conclude that the width of the waveguide should be of order of 14
$\mu$m.

\subsection{The geometry of the system}

Now we can describe the geometry of the system at hand as shown in
Fig.~\ref{fig1}. The waveguide created by two parallel Bragg mirrors
is placed in the ($x,z$)-plane. A strong control field consisting of
a superposition of counter-propagating waves with the frequency
$\omega_{c}$ and the wavevectors ${\bf k}_c\pm{\bf k}=(0,0,k_c)\pm (
k,0,0)$ ($k\ll k_c$), couples the excited state $|2\rangle$ and the
ground state $|1\rangle$. The excited state $|2\rangle$ is also
coupled to the low energy state $|3\rangle$ by a weak probe field of
the frequency $\omega_{p}$ and the wavevector ${\bf k}_p=(0,0,k_p)$
(i. e. it propagates along $z$-axis). We consider a planar waveguide
with the transverse (i.e. along $y$-axis) width small enough so that
only one transverse mode of the probe beam needs to be considered
(see Appendix~\ref{ap}). We concentrate on the $0-$th mode
designated as $s_0(y)$. Therefore the probe field can be presented
in the form
\begin{eqnarray}
\label{Ep} {\bf E}_p({\bf r},t)={\bf e}_p {\cal E}_p(x,z)
s_0(y)e^{i(k_pz-\omega_p t)}+{\rm c.c}
\end{eqnarray}
where, ${\bf e}_p$ is the polarization vector and ${\cal E}_p(x,z)$
is the slowly varying envelope of the probe field.

The frequency (amplitude) of the control field $\omega_c$
($E_c(\textbf{r},t)$) is much higher than that of the guided probe
beam $\omega_p$ ($E_p(\textbf{r},t)$).  This leads
to a the effective refractive index of the Bragg mirrors for the
control field in the claddings to be of order one (see Appendix~\ref{ap}), allowing us
neglecting its transverse distribution (we however emphasize that this design is introduced only for the sake of simplicity). The control field can be
presented in the form
\begin{eqnarray}
{\bf E}_c(\textbf{r},t)= 2{\bf e}_c {\cal E}_c(x,z) f(x)\cos(kx) e^{
i(k_c z-\omega_c t)}+{\rm c.c}.
\end{eqnarray}
Here, ${\bf e}_c$ stands for the polarization vector, the function $f(x)$ describes the slowly varying distribution
of the control field in the $x$ direction, and  ${\cal E}_c(x,z)$ is
the complex amplitude of the control field. In experiments, the desired space dependence of $f(x)$ can
be achieved with the help of beam masks.

The total electric-field can be written down as ${\bf
E}(\textbf{r},t)={\bf E}_p(\textbf{r},t)+{\bf E}_c(\textbf{r},t)$
and  considered classically. Large difference between the frequencies of
the probe and control fields ($\omega_c\gg\omega_p$) allows us to
split the equations for ${\bf E}_p$ and ${\bf E}_c$, while large
difference in the field amplitudes ($E_c\gg E_p$) allows us to
consider ${\bf E}_c$ to be constant in the equation for the probe
beam~\cite{FIM}. Now the equation governing ${\bf E}_p$ can be
written down as follows:
\be
\label{Max} \left(\nabla^2-\frac{1}{c^2}\frac{\partial^2 }{\partial
t^2}\right)  E_p=\frac{1}{\epsilon_0c^2}\frac{\partial^2 }{\partial
t^2} P.
\ee
Here $P$ is the probe beam polarization defined by the properties of
the atomic gas in the waveguide core and by the properties of the
Bragg mirrors in the waveguide claddings, i.e.
\be
P=\begin{cases}
P_{\rm clad}, \quad (y<-b, y>b)\\
P_{\rm core}.\quad (-b<y<b)
\end{cases}
\ee
Following the standard procedure~\cite{Heebner} $P_{\rm clad}$ can
be determined by calculating the effective refractive index of the
Bragg mirrors $n$ from Eq. (\ref{app_p}): $P_{\rm
clad}=\epsilon_0(n^2-1)E_p$.

\subsection{Nonlinear Polarization}

In order to compute $P_{\rm core}$ one introduces  the bosonic field
operators $\hat\psi_j$, $\hat\psi_j^\dag$ of the states $|j\rangle$
($j=1,2,3$), as well as the electric dipole matrix elements $p_{ij}$
associated with the transitions between the states $|i\rangle$ and
$|j\rangle$. Then the polarization can be obtained as
\begin{eqnarray} \label{p_core}
P_{\rm core}=p_{32}  \langle\hat{\psi}_3^{\dag} \hat{\psi}_2\rangle
e^{i( k_p x-\omega_pt)}+{\rm c}.{\rm c}.
\end{eqnarray}

The steering of the light we are interested in is performed through
the proper spatial and/or temporal modulations of $P_{\rm core}$. In
order to link it directly to the probe field $E_p$ we have to
address the Heisenberg equations for the operators $\hat\psi_j$
describing the atomic medium. This can be done neglecting the atomic
kinetic energy, what leads to the system as follows (its derivation repeats the
steps outlined say in~\cite{HKH})
\bes \label{Heisen}
\bea
& &  i\frac{\partial}{\partial
t}\hat{\psi}_1=-\Delta_1 \hat{\psi}_1-2\Omega_c^{\ast}f(x)\cos(k x)\hat{\psi}_2,\\
%& & \left(i\frac{\partial}{\partial
%t}-\Delta_2\right)\hat{\psi}_2=-2\Omega_c f(x)\cos(k x)\hat{\psi}_1
%\nonumber\\
%& & \quad\quad
%-\Omega_p\hat{\psi}_3,
& &  i\frac{\partial}{\partial
t}\hat{\psi}_2=(\Delta_2-i\gamma_2) \hat{\psi}_2-2\Omega_c f(x)\cos(k x)\hat{\psi}_1
%\nonumber\\
%& & \quad\quad
\nonumber\\
& & \quad\quad\quad\quad-\Omega_p\hat{\psi}_3,
\\
& & i\frac{\partial}{\partial t}\hat{\psi}_3=
-\Omega_p^{\ast}\hat{\psi}_2.
\label{Heisen_a}
\eea
\ees
Here $\Omega_{p}= p_{23}{\cal E}_p/\hbar$ and $\Omega_{c}=
p_{21}{\cal E}_c/\hbar$, are the Rabi frequencies of the respective
fields $\Delta_2=(\omega_2-\omega_3)-\omega_p$, and
$\Delta_1=(\omega_3-\omega_1)-(\omega_c-\omega_p)$ present  one-
and two-photon detunings, respectively (see Fig.~\ref{fig1}). In
Eqs.~(\ref{Heisen}), we  keep only the dissipation of the state $|2\rangle$ due to spontaneous emission, by adding the decay rate $\gamma_2$, phenomenologically.
Usually, the decay rates of the lower states $|1\rangle$ and
$|3\rangle$ are much smaller than that of the exited state and can be safely neglected. For a particular choice
of the atomic gas and assuming that the decay rate of the excited
state is much less than the two-photon detuning ($\gamma_2\ll\Delta_1$), one may consider
the loss of the atoms from the system as a small perturbation
(see also the discussion in~\cite{HKH}).

Since the two-photon detuning is nonzero (i.e. $\Delta_1\neq0$), the
pure dark state (corresponding to $\hat\psi_2\equiv 0$) can not exist. However, we explore the situation where the atomic
system is close enough to the dark state. To this end we look for
a solution of Eqs.~(\ref{Heisen}) in the form $\hpsi_{j}({\bf
r},t)=e^{i\lambda t}\hat{\phi}_j({\bf r})$, where the exponent
$\lambda$ is slowly varying in space and can be obtained as a root
of the respective characteristic equation. We are interested in a
root having the smallest imaginary part~\cite{note}, i.e.
 \be \label{lambda_r}
\lambda=\lambda_r+i\lambda_i, \ee
($|\lambda_i|\ll|\lambda_r|$) with
 \bea
& &\lambda_r=\frac{\Delta_1 |\Omega_p|^2}{\Delta_1
\Delta_2 +4|\Omega_c|^2f^2(x)\cos^2(k x)+|\Omega_p|^2},\nonumber\\
& &\lambda_i=\frac{\gamma_2}{ |\Omega_p|^2}\lambda_r^2.\nonumber
\eea
Then the expectation value $\langle\hat{\psi}_3^{\dag}
\hat{\psi}_2\rangle$ can be readily computed
from (\ref{Heisen}) and (\ref{lambda_r}):
\begin{equation}
\label{g_0}
%& &
\langle\hat{\psi}_3^{\dag} \hat{\psi}_2\rangle
=
%\nonumber\\
%& &
\frac{\Lambda
|\delta_1-\Lambda|^2\Omega}{(|\Lambda|^2+|\Omega|^2)|\delta_1-\Lambda|^2+4|\Lambda|^2f^2(x)\cos^2(k
x)}.
\end{equation}
Here we introduced the dimensionless functions $\Omega=\Omega_p/\Omega_c$ and
\bea \Lambda = \frac{\lambda}{|\Omega_c|}=\Lambda_r+i\Lambda_i, \eea
where \bea
\Lambda_r&=&\frac{\delta_1|\Omega|^2}{\delta_1\delta_2+4f^2(x)\cos^2(k
x)+|\Omega|^2}, \nonumber
\\
\Lambda_i&=&\frac{\tilde{\gamma}_2}{|\Omega|^2}\Lambda_r^2,
\nonumber \eea
as well as the parameters $\delta_j=\Delta_j/|\Omega_c|$ ($j=1,2$) and $\tilde{\gamma}_2=\gamma_2/|\Omega_c|$.
Since the probe field (decay rate of the excited state) is assumed to be weak   in comparison with the
control field, we have that $|\Omega|\ll 1$ ($\tilde{\gamma}_2\lesssim|\Omega|^2\ll 1$, see Eq. (\ref{g_1}) and the related discussions).

An important conclusion follows from Eq. (\ref{Heisen_a}), the
quasi-dark state, characterized by the small population of the
excited state $|2\rangle$, requires $|\lambda|\ll|\Omega_p|$ and
hence $|\Lambda|\ll 1$.

In what follows we will be interested in a particular choice
$f(x)=1- ax$ where $a$ is the mask parameter considered to be small.
More specifically, we will consider the beam propagation (i.e.
distributions of $\Omega$ in space) characterized by finite
deviations from the axis $x$. As we have already announced in the
Introduction, the mentioned deviation will occur due to BOs of the
beam. Thus if we assume that the amplitude of BOs measured in units
$k^{-1}$ [see Eq. (\ref{Max3}) below] is $X$, the smallness of $a$
is determined by the requirement that $a X/k\sim \Omega^2 \ll 1 $.
This allows us to achieve further simplification for the
polarization by defining
\be \label{v} v(k x)\equiv \delta_1\delta_2+4 \cos^2(k x).
\ee
Performing the expansion of the right hand side of Eq.
(\ref{g_0}) in the Taylor series with respect to $\Omega$ we get:
\bea \label{g_1} & & \langle\hat{\psi}_3^{\dag} \hat{\psi}_2\rangle
\simeq \frac{\delta_1\Omega}{v(k x)}
\left[1+\frac{8 a x \cos^2(k x)}{v(k x)} \right.\nonumber\\
& & \left. \quad\quad-\frac{\delta_1^2-\delta_1\delta_2+2v(k x)}{v(k
x)^2}|\Omega|^2\right]+i\frac{\delta_1^2\tilde{\gamma}_2}{v(k x)^2}\Omega. \eea
Here we have neglected the $\Omega^5$-order terms and assumed that in
the domain of BOs the condition $|\Omega|^2\ll v(x)$ is satisfied.
Notice that the requirement $v(x)\neq0$ holds for in the whole space
provided
\be \label{v_cond} \delta_1\delta_2>0 \quad{\rm or}\quad
\delta_1\delta_2<-4. \ee
This condition will be imposed in what follows.

\subsection{The equation for the beam envelope}

We are particularly interested in the beam dynamics resulting from the
interplay of diffraction and nonlinearity when the probe field
passes through the atomic cloud. Therefore, we concentrate on
stationary solutions, requiring $\Omega$ to be independent on time.
Substituting Eqs. (\ref{Ep}) and (\ref{g_1}) into Eq. (\ref{Max}),
taking into account (\ref{app_s}), and leaving only the leading
order terms, we arrive at the dimensionless equation for the slowly
varying function $\Omega$:
\bea
\label{Max3} & & i\frac{\partial\Omega}{\partial\zeta}+ \frac{\partial^2
\Omega}{\partial \xi^2} +U(\xi)\Omega +\xi F( \xi)\Omega
-G(\xi)|\Omega|^2\Omega\nonumber\\
& & \quad\quad+i\Gamma(\xi)\Omega=0.
\eea
Here  $\xi=kx$ and $\zeta=(k^2/2k_p)z$ are the dimensionless
independent variables,  $v(\xi)=v(k x)$ is given by (\ref{v}), and
we defined
\bea U(\xi)&=& \frac{U_0}{v( \xi)},\quad
F(\xi)= \frac{8 a^{\prime} U_0\cos^2(\xi)}{v^2(\xi)}, \nonumber\\
G(\xi)&=& \frac{C_0U_0[\delta_1^2-\delta_1\delta_2+2
v(\xi)]}{v^3(\xi) }, \quad
\Gamma(\xi)=\frac{\delta_1\tilde{\gamma}_2U_0}{v( \xi)^2},
\nonumber
\eea
with %
\be U_0=\frac{{\cal N}|{\bf
p}_{32}|^2k_p^2}{\hbar\epsilon_0k^2\Omega_c}\delta_1, \nonumber \ee
and $a^{\prime}=a/k$. The constant $C_0$ depending on the particular
choice of the Bragg mirrors is computed in the Appendix~\ref{ap}.

The obtained dimensionless Eq. (\ref{Max3}) implies that all the
terms are of the  order one or less. More specifically, it is
necessary to require that $U_0\sim 1 $. This last constrain can be
satisfied by different means. Notice that the parameter $k_p/k$ can
be experimentally controlled by the geometry of the laser beams (see
Fig.~\ref{fig1}), we thus fix $k_p/k=0.2$ and adjust other
parameters (in particular $\delta_1$ and $\Omega_c$) to satisfy the
constrain. For a typical set of data explored below, i.e. the atomic concentration ${\cal N}\approx10^{14}$ cm$^{-3}$ and
$\Delta_1=\Omega_c=1.0\times 10^7$ s$^{-1}$ ($\delta_1=1$), we
obtain $U_0=4.7$ satisfying the desired order of magnitude.
Meanwhile, the dissipation is very small ($\Gamma(\xi)\ll1$) in the
considered system since $\tilde{\gamma}_2=4.7\cdot10^{-3}\ll1$.

%%%%%%%%%%%%%%%%%%%%%%%%%%%%%%%%%%%%%%%%%%%%%%%%%%%%%%%%%%%%%%%%%
\section{Spatial gap solitons and all-optical steering}
\label{sec:III}

\subsection{Spatial gap solitons}
\label{sec:IIIa}

We study the situation when the effects of the linear force,
nonlinearity, and dissipation are small enough. Therefore we first
notice that for $F\equiv0$, $G\equiv0$, and $\Gamma\equiv0$ one
arrives at the familiar framework of the band theory, where by means
of substitution $\Omega\varpropto e^{-iK_\alpha\zeta}$
Eq.~(\ref{Max3}) is reduced to the eigenvalue problem
\be \label{Linear} \hat{{\cal L}}u_{\alpha,q}(\xi)+K_\alpha(q)u_{\alpha,q}(\xi)=0.
\ee
Here $\hat{{\cal L}}=-\partial^2/\partial \xi^2-U(\xi)$ and
$u_{\alpha,q}(\xi)$ is a Bloch function with indexes $\alpha$ and
$q$ standing respectively for the band index and for the wave vector
in the first Brillouin zone (BZ), i.e. for $q\in [-1,1]$ in the
case at hand. The physical meaning of the eigenvalue $K_\alpha(q)$
is the propagation constant. In what follows we concentrate on the
lowest energy band ($\alpha=0$) and therefore, for the sake of
brevity, omit the band index.

The effect of the  periodic linear
force originated by the control beam can be described by the
equations
\be \label{Motion} \frac{d \Xi}{d \zeta}=\tan\theta =\frac{d K_\alpha(Q)}{dQ}, \quad
  \frac{d Q}{d \zeta}=-F(\Xi), \ee
where $\Xi$ and $Q$ denote the coordinates of the center of the
Bloch wave packet in the real and reciprocal spaces, respectively.
$\theta$ is the refraction angle with
$\tan\theta=2(k_p/k)\tan\theta_p$ (see Figs.~\ref{fig1} (a) and
~\ref{fig3} (a)). A peculiarity
of our case is that $F(\xi)$ is an oscillating function, which has a
nonzero mean value
$$\langle
F\rangle=\frac{1}{\pi}\int_{0}^{\pi}F(\xi)=\frac{
4a^{\prime}U_0}{|\delta_1\delta_2+4|\sqrt{(\delta_1\delta_2+4)\delta_1\delta_2}},$$
and hence results in the oscillation behavior of the Bloch wave
packet in the real and reciprocal spaces (i.e. in BOs). The
dimensionless amplitude (along the $x$ axis) and period (along the
$z$ axis) of such oscillations can be calculated as $X=\Delta
K/2\langle F\rangle$, with $\Delta K$ denoting the width of the
lowest allowed band of the periodic potential $U(\xi)$, and
$Z=2/|\langle F\rangle|$, respectively.

In order to describe the effect of the nonlinearity we employ
the standard multiple-scale expansion assuming smallness of the
averaged force, i.e. $\langle F\rangle\ll1$. This last requirement
is equivalent to the condition $a\ll k$, which is consistent with
the constrains on the BOs amplitude discussed in Sec. II C provided $X\lesssim 1$.

Next we introduce the scaled variables
$(\xi_j,\zeta_j)\equiv\mu^{j/2}(\xi,\zeta)$ ($j=1$, 2,  $\cdots$), where
$\mu$ is a small parameter estimated by $\mu \sim \langle
F\rangle\ll 1$. The probe field can be written in the form of the
expansion
$\Omega = \sum_{j=1}^\infty\mu^{j/2}\Omega^{(j)}(\xi_0,\zeta_1),$
where in the arguments of the functions $\Omega^{(j)}$ we have
indicated only the most rapid variables. $\Gamma(\xi)=\mu^{2}{\cal R}( \xi_0)$. Substituting this ansatz
into Eq. (\ref{Max3}) and collecting the terms at each order of
$\mu^{1/2}$, we obtain a set of equations which can be solved order
by order (see e.g.~\cite{KS}). Omitting the details we turn directly
to the equation for the slowly varying amplitude $A(\chi,\zeta_2)$,
which is defined through the relation
$\Omega^{(1)}=A(\chi,\zeta_2)e^{iK(Q)\zeta_0}u_{Q}(\xi_0)$, where
$\chi=\xi_1-\tan\left[\theta(\zeta_2)\right]\zeta_1$ and the
dependence $Q=Q(\zeta_2)$ and $\theta(\zeta_2)$ are obtained from
Eqs. (\ref{Motion}). In the third order of the asymptotic expansion,
${\cal O}(\mu^{3/2})$, we obtain
\be\label{NLS} i\frac{\partial A}{\partial \zeta_2}+{\cal
D}(Q)\frac{\partial^2 A}{\partial \chi^2}-{\cal G}(Q)|A|^2A+i{\cal R}(Q)A=0, \ee
where the coefficients of the diffraction ${\cal D}$ and the effective
nonlinearity ${\cal G}$ are given respectively by
\bea
& &
{\cal D}(Q)=\frac 12 \frac{d^2 K(Q)}{d Q^2},
\quad
%\nonumber\\
%& &
{\cal G}(Q)=\int_{0}^{\pi} G( \xi)|u_{Q}(\xi)|^4 d\xi, \nonumber\\
& &{\cal R}(Q)=\int_{0}^{\pi} {\cal R}( \xi)|u_{Q}(\xi)|^2 d\xi. \nonumber
\eea

We notice that both ${\cal D}$ and ${\cal G}$ can be either
positive or negative  depending on the signs of detunings and  on the position of the center of the wave packet (i.e. on $\zeta_2$).  Therefore, in a general case BOs
oscillations will be accompanied by significant change of the width
of the beam, due to the diffraction. This undesired effect, however
can be dramatically reduced, if the conditions for self-focusing of
the beam will be satisfied (and hence will compensate the diffraction) at any coordinate $\zeta_2$. These
requirements (which are also referred to  as  conditions for the gap
soliton existence or, alternatively, as conditions of the modulational instability of
the Bloch waves), are well known and read~\cite{KS,condition}
\begin{eqnarray}
\label{main_cond}
{\cal D}(Q){\cal G}(Q)<0.
\end{eqnarray}
If condition (\ref{main_cond}) is satisfied for all (or almost all) wave-vectors
$Q$, the self-focusing of the beam will occur along the whole (or almost whole)
trajectory of the beam.

%===========================fig2===============================%
\begin{figure}
\centering
\includegraphics[scale=0.4]{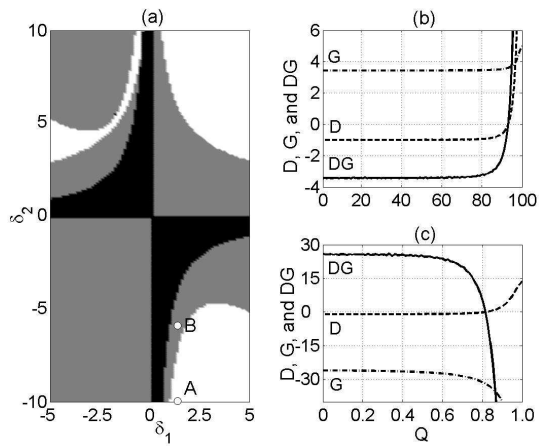}
\caption{(a) The white domains correspond to the choice of the
detuning, such that (\ref{main_cond}) is valid for 90\% of
wavevectors in the first BZ; in the black domains the requirement
(\ref{v_cond}) is not satisfied and thus the approximation
Eq.~(\ref{Max3}) fails; the gray domains show the parameter region
where our approximation is valid, but the conditions
(\ref{main_cond}) are not satisfied for more than 10\% of the
wavevectors in the BZ. The system parameters are given in Sec. II A.
  (b) The coefficients ${\cal D}(Q)$ (dashed line), ${\cal
G}(Q)$ (dash-dotted line), and ${\cal D}(Q){\cal G}(Q)$ (solid line)
{\it vs} $Q$ with the parameters $U_0=4.7$, $\Delta_1=\Omega_c=1.0\times 10^7$
s$^{-1}$, and $\Delta_2=-1.0\times 10^8$ s$^{-1}$ [the point $A$ in panel
(a)]. (c) The same as in panel (b)  but with $\Delta_2=-0.6\times 10^8$ s$^{-1}$
 [the point $B$ in panel (a)]. The
dynamics of the Bloch wave packets for the points $A$ and $B$ is shown
below in the panels  (a) and (b) of Fig.~\ref{fig3}. } \label{fig2}
\end{figure}
%===========================fig2===============================%
Thus, as the third step, for achieving the best performance of the
device one has to design the system (geometry of the beams or photon
detunings) in a way  to ensure the condition (\ref{main_cond}) for
all $Q$. The procedure of performing this task was described in
the previous works~\cite{condition,SKB} and here we report only the
final result. More specifically, in Fig.~\ref{fig2} (a), we show the
domains on the plain ($\delta_1,\delta_2$) where the optimal
conditions are satisfied (empty regions), not satisfied (gray
regions), and the parameter requirement (\ref{v_cond}) is violated (black
regions). In Fig.~\ref{fig2} (b) and (c), we plot the curves ${\cal
D}(Q)$, ${\cal G}(Q)$, and ${\cal D}(Q){\cal G}(Q)$ versus $Q$ with
the almost optimal design [point $A$ in panel (a)] and non optimal
design [point $B$ in panel (a)] of the OLs. We  mention that
always there exists a small range of the wavectors where spatial gap
solitons do not exist (i.e.  where ${\cal D}(Q){\cal G}(Q)>0$). This
range, however, can be made enough small [less than $10\%$ of the
whole BZ, like this happens in the point $A$, see Fig.~\ref{fig2} (b)], and therefore  has no appreciable
destructive effect on the beam dynamics [see Fig.~\ref{fig3} (a)].
On the contrary, the point $B$ in panel (a) corresponds to the case in
which the BOs of the beam are rapidly destroyed leading to strong
defocussing of the outcome beam [see Fig.~\ref{fig3} (b)].

\subsection{Long-living BOs and all-optical steering}

\begin{figure}
\begin{center}
\includegraphics[scale=0.45]{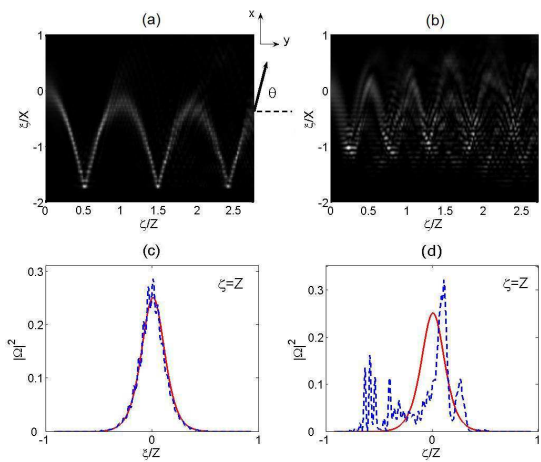}
\end{center}
\caption{(color on line) (a) The long-living BOs of the beam in the
almost optimally designed OLs [point $A$ in Fig.~\ref{fig2} (a)] obtained by direct numerical integration of Eq.~(\ref{Max3}).
The amplitude and the period of the  BOs are $X=14.47$ ($=49$
$\mu$m) and $Z=578.98$ ($=787$ $\mu$m). (b)  BOs of the wave packet
in the non optimally designed OLs [point $B$ in Fig.~\ref{fig2}
(a)]. In the both cases the initial condition is a stationary gap
soliton located at $\xi/X=0$. The intensity profiles of the output
beam (blue, dashed line) vs the input beam (red, solid line)
corresponding to the cases (a) and (b) are respectively shown in the panels  (c)
and (d)  obtained at $\zeta=Z$.
} \label{fig3}
\end{figure}

Now we turn to the final step of numerical descriptions of the
   BOs of a spatial soliton and of the ight steering. In Fig.~\ref{fig3} (a),
we show the dynamics of the Bloch wave packet launched normally in
the atomic media with the almost optimal design of the OLs [point
$A$ in Fig.~\ref{fig2} (a)]. The result was obtained by employing
numerical integration of Eq.~(\ref{Max3}). The existence of the
long-living BOs of the gap soliton, with the amplitude and period
well matching the theoretical estimates, can be observed. Due to the
very small dissipation, there is only a tiny decrease of the soliton
intensity due to the absorption of the medium. More specifically,
one can see the highly concentrated beam in the output with the
direction of propagation determined by the nonzero angle $\theta$.
For the sake of comparison, in Fig.~\ref{fig3} (b), we show the
dynamics of the same Bloch wave packet in the non optimally designed
OL [The parameters used in this panel are the same with those used
in Fig.~\ref{fig2} (c)]. In this case, the beam undergoes strong
spreading out due to the alternating  diffractive and self-defocussing regeems. The
intensity profiles of the output beam vs the input beam
%corresponding to the cases (a) and (b)
are
%respectively
shown in
the panels (c) and (d). One observes that there is no apparent deformation
between the output beam and input beam in panel (c) while an obvious
destruction appears in panel (d).
%===========================fig3===============================%
%===========================fig3===============================%

The maximum input intensity of the probe beam in Fig.~\ref{fig3} (a)
can be estimated by the formula $I_{\rm max}=(c/2)\epsilon_0|{\bf
E}_{p\,{\rm max}}|^2$. Using the parameters given in Fig.~\ref{fig2}
(b), we obtain $I_{\rm max}\approx4.5$ $\mu$W cm$^{-2}$. Thus, the
generation  of the spatial gap soliton in the system at hand
requires only very low input light intensity. We remark that the
intensity of a single 500-nm photon per nanosecond on an area of 1
$\mu$m$^2$ is $I_{ph}=0.04$ W cm$^{-2}$. This estimation shows that
our system makes it possible to manage the weak beams characterized
by single-photon wavepackets. This is drastically different from the
spatial optical soliton generation in a conventional waveguide where
an input laser pulse with very high peak power $\sim10^2$ kW is
needed in order to bring out a sufficient nonlinear
effect~\cite{Aitchison}.

\begin{figure}
\begin{center}
\includegraphics[scale=0.45]{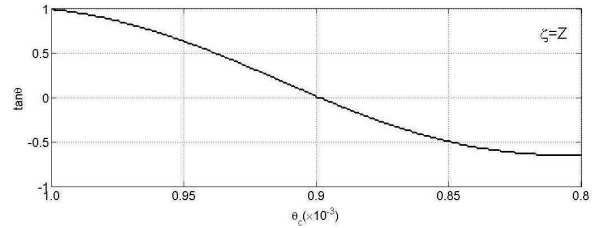}
\end{center}
\caption{$\tan\theta$ {\it vs} $\theta_c$ at $\zeta=Z$ for the
almost optimally designed OLs [corresponding to point $A$ in Fig.~\ref{fig2} (a) and panel (a) of Fig.~\ref{fig3}].
%:
%$\Delta_1=\Omega_c=1.0\times 10^7$ s$^{-1}$, $\Delta_2=-1.0\times
%10^8$ s$^{-1}$].
} \label{fig4}
\end{figure}

The existence of long-living BOs of
spatial optical gap solitons can be used to implement efficient
all-optical steering of  probe beams. Different from all linear
systems where the spread and attenuation of the probe pulse are
unavoidable because the refractive index gradient depends on both
light frequency and spatial coordinates, the scheme proposed here
can greatly improve the device performance due to the
self-focusing of the beam. In experiments, the optical masks and
atomic cells are usually chosen in advance. However, one can exploit the possibility of changing
the geometry or parameters of the control field to steer the output
probe beam. As an example, by changing the angle between the input
control fields $\theta_c=\arctan(k/k_c)$ (see Fig.~\ref{fig1}), one
can efficiently control the refraction angle of the output probe
beam. In Fig.~\ref{fig4}, we show the tangent of the refraction
angle $\tan\theta$ versus $\theta_c$ with a particular length of
atomic cells (i.e. $\zeta=Z$). A wide range of angle,
$\theta_p\in[-0.98,1.19]$, can be achieved with the given
parameters.

%===========================fig4===============================%
%===========================fig4===============================%

%%%%%%%%%%%%%%%%%%%%%%%%%%%%%%%%%%%%%%%%%%%%%%%%%%%%%%%%%%%%%%%%%%
\section{Conclusion}

In conclusion, we studied (both analytically and numerically) the
dynamics of weak probe optical beams in a gas of three-level
$\Lambda$-atoms subjected to a control laser beam. The standing
wave originates periodic modulations of the linear and nonlinear
parts of the dielectric permittivity of the medium, which is
``seen'' by the probe beam. If the amplitude of the
control field is smoothly modulated, the probe beam undergoes the
Bloch oscillations. These oscillations correspond to periodic change
of the direction of the propagation of the beam. We have shown that it is possible to
choose the parameters of the atomic system, say Rabi frequency of the control field and photon detunings, to reduce the diffraction of the probe beam,
and thus the output beam is approximately the same with
the input one. In this case, the probe field distributions is
nothing but a spatial gap soliton undergoing long-living Bloch oscillations.
Due to the significant enhancement of nonlinearity, such
solitons can be formed with extremely weak light intensity, below
single-photon wavepacket level. Thus, through changing the
geometry or parameters of the system one can control the
direction of the output beam to carry out an efficient all-optical steering of light.

%%%%%%%%%%%%%

\acknowledgments

The work of C.H. was supported by the Funda\c{c}\~ao para a
Ci\^encia e a Tecnologia (FCT) under Grant No. SFRH/BPD/36385/2007
and Est\'{i}mulo \`{a} Investiga\c{c}\~{a}o 2009 de Funda\c{c}\~{a}o
Calouste Gulbenkian. The research of VVK was partially supported by
the grant PIIF-GA-2009-236099 (NOMATOS).

\appendix

\section{The transverse profiles of the probe field}
\label{ap}

The transverse profiles of the probe field can be determined from
the eigenvalue problem~\cite{Heebner}
\bes \label{app_s}
\bea & &\left(\frac{\partial^2 }{\partial y^2}+\frac{\omega^2}{c^2}-k_{l}^2\right)s_l(y)=0,\, (|y|<b) \label{app_s1}\\
& &\left(\frac{\partial^2 }{\partial
y^2}+\frac{n^2\omega^2}{c^2}-k_{l}^2\right)s_l(y)=0,\, (|y| >b)
\label{app_s2} \eea \ees
where $\omega$ stands for either $\omega_p$ or $\omega_c$,  $l$
stands for the mode index ($l=0,1,...$), $s_l(y)$ is a normalized,
i.e. $\int_{-\infty}^{\infty}dy|s_l(y)|^2=1$, profile of the
respective mode, and $2b$ is the width of the waveguide core.

The eigenvalue $k_{l}$ and effective refractive index of the Bragg
mirrors $n$ can be obtained from the dispersion relations of the
planar waveguide
\be \label{app_w}
\tan^2\left[b\sqrt{\frac{\omega ^2}{c^2}-k_{l}^2}-\frac{(l+1)\pi}{2}\right]=\frac{k_{l}^2c^2-\omega ^2}{n^2\omega^2-k_{l}^2c^2},
\ee
 and the expression valid the finite photonic
structures~\cite{Centini}
\bea \label{app_p} n(\omega)=\frac{c}{\omega
h}\left(\phi_t-\frac{i}{2}\ln T(\omega)\right), \eea
where $\phi_t(\omega)$ and $T(\omega)$ are the total phase and the transmittance of the complex transmission coefficient for the structure
$t(\omega)=x(\omega)+iy(\omega)=\sqrt{T}e^{i\phi_t}$
with $\phi_t=\arctan(y/x)\pm m\pi$ (the integer $m$ is uniquely
defined assuming $\phi_t$ is a monotonically increasing function) and $T(\omega)=x^2+y^2$.
$h$ is the total thickness of the $N$-period crystal $h=N(h_a+h_b)$
with $h_a$ and $h_b$ being the thicknesses of two different
materials [see Fig.~\ref{fig1} (b)].

As an example we consider  the probe field  constrained of
 quarter-wave Bragg mirrors, $n_ah_a=n_bh_b=\lambda_p/4$
with $n_a=1.0$ and $n_b=1.4$ being the refractive indexes of the two
different materials, for claddings, $N=2$, and $2b=0.1$ mm, we
obtain that $k_{0}=1081$ cm$^{-1}$ and $n\approx 0.87$. At the same time, for the control field we have
$k_{0}\approx 91144$ cm$^{-1}$  and the effective refractive index
of the cladding is $n\approx 1.0$.

The assumption that only $s_0$ takes into account requires that the
energy for exciting $s_1$ [corresponding to the eigenvalue of Eq.
(\ref{app_s1})] is higher than that of the self-action
(nonlinearity) in Eq.~(\ref{Max3}). With the parameters given in
Fig.~\ref{fig2}, we obtain
$G(\xi)|\Omega|^2\Omega<(\omega_p^2/c^2-k_{1}^2)/k^2$ ($k_{1}=873$
cm$^{-1}$ for the probe field), hence we can safely neglect the
higher transverse modes (i.e. $s_l$ for $l>0$). Here, the integral
$C_0=\int_{-\infty}^{\infty}|s_0|^4d\eta=73.9$. However, the much
higher frequency of the control field allows us to neglect the
difference among transverse profiles of the control field in the
waveguide core.

%%%%%%%%%%%%%%%%%%%%%%%%%%%%%%%%%%%%%%%%%%%%%%%%%%%%%%%%%%%%%%

%%%%%%%%%%%%%%%%%%%%%%%%%%%%%%%%%%%%%%%%%%%%%%%%%%%%%%%%%%%%%%


\begin{references}
\bibitem{note} The two other roots describe fast decay of the respective expectation values  and therefore  not
  considered here.
\end{references}
\end{document}